\documentclass[twocolumn,english,superscriptaddress,prb,longbibliography,nobibnotes, groupedaddress]{revtex4-1}
\usepackage{xcolor}
\usepackage{bm}
\usepackage{amstext}
\usepackage{amssymb}
\usepackage{graphicx}
\usepackage{commath}
\usepackage{dsfont}
\usepackage{mathtools}
\usepackage{amsmath,mathrsfs}
\usepackage{physics}
\usepackage{braket}
\usepackage{hyperref}
\usepackage{algpseudocode}
\usepackage{algorithm}
\usepackage{comment}
\usepackage{soul}

\newcommand{\sm}{\color{red}}

\begin{document}
\title{Universal Counterdiabatic Driving in Krylov Space}

\author{Stewart Morawetz}
\email{morawetz@bu.edu}
\affiliation{Department of Physics, Boston University, Boston, Massachusetts 02215, USA}

\author{Anatoli Polkovnikov}
\affiliation{Department of Physics, Boston University, Boston, Massachusetts 02215, USA}

\begin{abstract}

Local counterdiabatic (CD) driving provides a systematic way of constructing a control protocol to approximately suppress the excitations resulting from changing some parameter(s) of a quantum system at a finite rate. However, designing CD protocols typically requires knowledge of the original Hamiltonian \textit{a priori}. In this work, we design local CD driving protocols in Krylov space using only the characteristic local time scales of the system set by e.g. phonon frequencies in materials or Rabi frequencies in superconducting qubit arrays. Surprisingly, we find that convergence of  these universal protocols is controlled by the asymptotic high frequency tails of the response functions. This finding hints at a deep connection between the long-time, low frequency response of the system controlling non-adiabatic effects, and the high-frequency response determined by the short-time operator growth and the Krylov complexity. We make this connection concrete by showing how, for a representative integrable model, we may extract long-time universal behaviour of the correlation functions from a short-time expansion of the dynamics using a system-independent universal protocol.

\end{abstract}

\maketitle

\section{Introduction} \label{sec:introduction}

Recent years have seen rapid advances in experimental platforms such as ultracold atoms \cite{aidelsburgerRealizationHofstadterHamiltonian2013, grossQuantumSimulationsUltracold2017, ebadiQuantumPhasesMatter2021}, trapped ions \cite{friisObservationEntangledStates2018, zhangObservationManybodyDynamical2017} and superconducting qubits \cite{barendsDigitizedAdiabaticQuantum2016, maDissipativelyStabilizedMott2019, eickbuschFastUniversalControl2022,  googleSuppressingQuantumErrors2023, kingQuantumCriticalDynamics2023} which enable the precise simulation of closed, interacting quantum systems and provide promising platforms for quantum computing \cite{henrietQuantumComputingNeutral2020, haffnerQuantumComputingTrapped2008, kjaergaardSuperconductingQubitsCurrent2020}. These motivate the need for precise control over the quantum states that may be realized in these experiments. One such technique is adiabatic processes \cite{albashAdiabaticQuantumComputation2018}, wherein a parameter of the system (e.g. laser frequency, magnetic field strength, etc.) is changed slowly enough that excitations resulting from this drive are suppressed. Then, if the system begins in an eigenstate, it will remain in an instantaneous eigenstate of the system at all times. Adiabatic processes are important in many different contexts such as quantum annealing, quantum state preparation, and various thermodynamic processes because they are reversible and therefore do not have uncontrolled dissipation that leads to the generation of various errors or energy losses.

The main deficiency of adiabatic processes is that they generally require very long time scales. This fact severely limits, for example, the power of thermal engines operating at Carnot efficiency or equally the operation times of computing algorithms based on quantum annealing. Even for the finite-sized systems currently accessible in modern experiments, the timescales required in order to remain adiabatic are often longer than the coherence times so that adiabatic processes are not even possible to realize. One approach to accelerate protocols while staying adiabatic are so called ``Shortcuts to Adiabaticity" (STA)~\cite{Guery-Odelin2019}. Then, the adiabatic evolution of a system can be realized in finite time, at the expense of adding extra control terms to the driving protocol.

There are many techniques that fall under this umbrella, and in this work we will focus in particular on \textit{counterdiabatic (CD) driving}, which has received extensive theoretical \cite{demirplakAdiabaticPopulationTransfer2003, demirplakAssistedAdiabaticPassage2005, berryTransitionlessQuantumDriving2009, delcampoShortcutsAdiabaticityCounterdiabatic2013, selsMinimizingIrreversibleLosses2017, claeysFloquetengineeringCounterdiabaticProtocols2019, gjonbalajCounterdiabaticDrivingClassical2022, cepaiteCounterdiabaticOptimizedLocal2023, schindlerCounterdiabaticDrivingPeriodically2024, duncanExactCounterdiabaticDriving2024, morawetzEfficientPathsLocal2024, vanvreumingenGatebasedCounterdiabaticDriving2024, gjonbalajShortcutsAnalogPreparation2025} and experimental \cite{duExperimentalRealizationStimulated2016, anShortcutsAdiabaticityCounterdiabatic2016, huExperimentalImplementationGeneralized2018, boyersFloquetengineeredQuantumState2019} attention. This approach modifies the Hamiltonian of a system by adding to it a term which exactly cancels the part responsible for excitations when a parameter is changed. This enables adiabatic evolution without the typical requirement of long coherence times. We note that any STA protocol can be exactly mapped to an appropriate CD protocol~\cite{Bukov_2019} by some unitary rotation. This rotation can generally alter the path followed by the instantaneous state compared to the adiabatic path as well as the parent Hamiltonian for which this state is an eigenstate. Generally all STA protocols realize CD driving for some suitably deformed Hamiltonian (see also Refs.~\onlinecite{cepaiteCounterdiabaticOptimizedLocal2023,morawetzEfficientPathsLocal2024}).

Engineering CD protocols requires implementation of the Adiabatic Gauge Potential (AGP)~\cite{selsMinimizingIrreversibleLosses2017}, which is generally unknown for interacting systems. Moreover for chaotic systems it is even poorly defined being a very nonlocal operator with the squared norm scaling linearly with the Hilbert space-dimension or even faster~\cite{Pandey_2020}. The AGP is usually better defined if it only suppresses non-adiabatic transitions to/from the ground state, but even then it is generally a highly nonlocal operator, especially if the spectrum is gapless. Therefore finding the CD protocol exactly, let alone implementing it, is impractical beyond small systems. This has motivated the development of a variational approach: \textit{local counterdiabatic driving} \cite{selsMinimizingIrreversibleLosses2017}, where the suppression of excitations is not exact, and some error is introduced. This error can be systematically reduced by increasing the variational manifold of extra control terms. The variational principle does not rely on knowledge of the eigenstates of the instantaneous Hamiltonian it has to follow, but does need precise knowledge of this Hamiltonian. A particularly convenient variational ansatz can be obtained using operators from the so-called Krylov space~\cite{claeysFloquetengineeringCounterdiabaticProtocols2019,bhattacharjeeLanczosApproachAdiabatic2023,takahashiShortcutsAdiabaticityKrylov2024}.

In practice the Hamiltonian might not be known exactly due to presence of some uncontrolled terms. In this work, we ask whether one can find good ``universal'' CD protocols, which do not require any detailed information about the Hamiltonian of the system except for overall local energy scales, which set the natural time units. We find that it is generally possible to define such a universal CD protocol, but surprisingly its complexity is controlled not by the low-energy excitations (small gaps) as one might guess, but rather the high energy states which are artificially excited by the error in the approximate drive deviating from the true CD protocol. This result has potential implications beyond counterdiabatic driving, hinting at a connection between the short- and long-time dynamics of Hamiltonian systems. Note that while we focus on quantum systems here, a similar analysis equally applies to classical systems.

The paper is organized as follows: In Section \ref{sec:approximating_AGP} we review CD driving and discuss how finding variational CD protocols in Krylov space can be cast as the mathematical problem of approximating $1/x$ by odd polynomials. Then we discuss how we can use this to find universal protocols, and how the high-frequency response of a system can limit the convergence of this procedure. In Section \ref{sec:examples} we give examples and numerically confirm our quantitative predictions about the effect of the high frequency response on convergence. We make this short- and long-time dynamics connection more explicit in Section \ref{sec:long-short-time-connection}, where we demonstrate how the universal protocol can be combined with a short-time expansion to extract universal behaviour in long-time correlation function of physical observables, or equivalently the low frequency asymptote of their spectral functions. Finally, we conclude in Section \ref{sec:conclusions}, highlighting the main results and suggesting possible future directions of exploration.

\section{Approximate Local CD Driving}
\label{sec:approximating_AGP}

A standard adiabatic approach to state preparation is \textit{quantum annealing}. Consider a Hamiltonian $H(\lambda)$ with some  parameter $\lambda \in [0, 1]$ which may be controlled. If the system begins in the ground state, the adiabatic theorem \cite{Sakurai2017} guarantees that so long as $H(\lambda)$ has finite gaps between energy levels, one can always change $\lambda$ sufficiently slowly that the probability of excitation out of the ground state is arbitrarily small. Then, one implements a Hamiltonian so that $H(\lambda = 0)$ has a ground state which is easy to prepare, and the ground state of $H(\lambda = 1)$ is the desired state.

Formally, this approach works because the terms coupling different states in the eigenbasis are proportional to the annealing rate $\dot{\lambda}$, so that making this rate sufficiently small will suppress any transitions. The approach of counterdiabatic driving is to modify the Hamiltonian so that these terms vanish exactly. If this is done, no transitions will occur even if $\dot{\lambda}$ is large. Mathematically, we define the \textit{counterdiabatic Hamiltonian} as

\begin{equation} \label{eqn:H_CD}
    H_{CD}(\lambda) = H(\lambda) + \dot{\lambda} A_\lambda,    
\end{equation}

\noindent where $A_\lambda$ is the so-called adiabatic gauge potential (AGP). The extra counter term in the Hamiltonian suppresses non-adiabatic transitions which lead to excitations from the ground state for any rate $\dot\lambda$. For a non-degenerate spectrum the matrix elements of this operator are (see Eq. (63) of [\onlinecite{kolodrubetzGeometryNonadiabaticResponse2017}])
\begin{equation} \label{eqn:exact-AGP-mtx-elts}
    \langle m \vert A_\lambda \vert n \rangle =-
    \lim_{\mu\to 0}i \frac{\omega_{mn}}{\omega_{mn}^2+\mu^2} \langle m \vert \partial_\lambda H \vert n \rangle,
\end{equation}
where $\omega_{mn}=E_m-E_n$, we have set $\hbar=1$, where we have inserted a regularization $\mu$ so that the matrix elements remains well defined as $\omega_{mn} \rightarrow 0$. Because this the regularization plays little role for $\omega_{mn} \gtrsim \mu$, doing CD driving with this AGP suppresses transitions between states $m \leftrightarrow n$ with $|E_n-E_m| \lesssim \mu$. Another way to think of $\mu$ is as the (inverse) time cutoff, above which all non-equal time correlation functions are truncated. In particular, if we are only interested in suppressing transitions from the ground state, we can set $\mu$ to be the gap between the ground and first excited states, which is usually significantly (exponentially) larger than the spectral gap in the middle of the spectrum. If this were realized, ground states could be prepared adiabatically with arbitrarily high fidelity.

The natural measure of complexity of the CD driving, which can be also termed as adiabatic complexity~\cite{lim2024definingclassicalquantumchaos}, is the averaged fidelity susceptibility~\cite{kolodrubetzGeometryNonadiabaticResponse2017}, defined as the norm of the AGP:
\begin{equation}
\label{eq:chi_lambda}
\chi_\lambda=\|A_\lambda\|^2=\lim_{\mu\to 0}{1\over 4\pi}\int_{-\infty}^{\infty} d\omega {\omega^2\over (\omega^2+\mu^2)^2}\Phi_\lambda(\omega),
\end{equation}
where
\begin{equation} \label{eqn:spectral-fn}
    \Phi_\lambda(\omega) = \sum_{m,n} \rho_n \vert \langle m \vert \partial_\lambda H \vert n \rangle \vert^2 \delta(\omega_{mn} - \omega)
\end{equation}
is the spectral function. Here the probabilities $\rho_n$ represent some stationary statistical ensemble, for example the Gibbs distribution at finite temperature. If we are interested in the ground state, then we can choose $\rho_n=\delta_{n0}$. Another convenient choice is the infinite temperature ensemble $\rho_n=1/\mathcal D$, with $\mathcal D$ the Hilbert space dimension. In this case $\chi_\lambda$ coincides with the square of the standard Frobenius norm of the AGP. From Eq.~\eqref{eqn:exact-AGP-mtx-elts} we see that the fidelity susceptibility is determined by the nearby energy states or equivalently according to  Eq.~\eqref{eq:chi_lambda} by the low-frequency scaling of the spectral function. In particular, for the ground state $\rho_n=\delta_{n0}$ 
\[
\chi_\lambda\leq \|\partial_\lambda H\|^2/\Delta^2\;\rightarrow \|A_\lambda\|\leq \|\partial_\lambda H\|/\Delta,
\]
where $\Delta=E_1-E_0$ is the energy gap. This result tells us that the complexity of the CD protocol is dictated by the minimal gap along the adiabatic path, as expected.  The same gap also sets the minimal annealing time in a standard adiabatic protocol and hence determines the adiabatic complexity in time. 

As already mentioned, generally the AGP is a highly non-local operator, therefore in order to make some progress in complex systems we must use some local approximation. A stable and robust approach to find such an approximate local AGP comes from a variational principle \cite{selsMinimizingIrreversibleLosses2017}, which minimizes the following action~\footnote{It is straightforward to check that extra $\mu^2 \|X\|^2$ term in the action leads to the same regularization of the AGP as in Eq.~\eqref{eqn:exact-AGP-mtx-elts}}:
\begin{equation} \label{eqn:variational_action}
    S_\lambda(X) = \|  G_\lambda(X)\|^2+\mu^2\| X\|^2;\;
    G_\lambda(X) = \partial_\lambda H + i [X, H].
\end{equation}
The AGP is defined as the exact minimum of the action, where, in the limit $\mu\to 0$, $[G_\lambda(A_\lambda),H]=0$. The approximate AGP can be found by minimizing the action over some space of local accessible operators.

It has been observed that a natural variational basis in which to construct the approximate AGP is to make an expansion in Krylov space \cite{claeysFloquetengineeringCounterdiabaticProtocols2019, Sels_2023, takahashiShortcutsAdiabaticityKrylov2024, bhattacharjeeLanczosApproachAdiabatic2023}, which corresponds to choosing a series of operators generated by the action of $[H, \boldsymbol{\cdot} \ ]$ on $\partial_\lambda H$ as the variational basis:
\begin{equation} \label{eqn:commutator_ansatz}
    A_\lambda^{(\ell)} = i \sum_{k = 1}^\ell \alpha_k \underbrace{[H,[H, ..., [H}_{2k-1}, \partial_\lambda H]]].
\end{equation}
The Krylov space is constructed by a proper Gram-Schmidt orthogonalization procedure in the space of nested commutators. One determines the $\alpha_k$ by substituting $X = A_\lambda^{(\ell)}$ into Eq. \eqref{eqn:variational_action} and choosing the $\alpha_k$ to minimize the action. Although these commutators may appear to be out of reach in a realistic experimental setting, they may be realized via the techniques of Floquet engineering \cite{bukovUniversalHighfrequencyBehavior2015, goldmanPeriodicallyDrivenQuantum2014, choiRobustDynamicHamiltonian2020} in which one does not need to couple to any terms not present in the Hamiltonian. Instead, rapid oscillation of the original couplings at high frequency can reproduce the commutators in an effective Floquet Hamiltonian.

\subsection{Variational AGP as polynomial fitting} \label{subsec:variational-poly-fitting}

The ansatz of Eq. \eqref{eqn:commutator_ansatz} leads to an interesting observation~\cite{claeysFloquetengineeringCounterdiabaticProtocols2019}. Start by writing the matrix elements of $A_\lambda^{(\ell)}$ in the basis of instantaneous eigenstates of $H(\lambda)$:

\begin{equation} \label{eqn:approx-AGP-mtx-elts}
    \langle m \vert A_\lambda^{(\ell)} \vert n \rangle = i \sum_{k=1}^\ell \alpha_k \omega_{mn}^{2k-1} \langle m \vert \partial_\lambda H \vert n \rangle,
\end{equation}

\noindent where we have suppressed the dependence of the eigenstates and energies on $\lambda$. Then, we compare this to the matrix elements of the exact AGP in Eq. \eqref{eqn:exact-AGP-mtx-elts}. The two agree if for every $\{m,n\}$~\footnote{If we are only interested in the ground state CD we can set $n=0$} we have:

\begin{equation} \label{eqn:equate-approx-exact-AGP}
    \left(\sum_{k=1}^\ell \alpha_k \omega_{mn}^{2k-1} + {1\over \omega_{mn}} g(\omega_{mn}/\mu) \right) \langle m \vert \partial_\lambda H \vert n \rangle  = 0,
\end{equation}
where $g(x)$ is the filter function satisfying $g(x\gg 1)=1$ and $g(x\to 0)\to 0$. For the regularization scheme used in Eq.~\eqref{eqn:exact-AGP-mtx-elts} this function is given by $g(x)=x^2/(1+x^2)$. This function is necessary to make the solution to Eq.~\eqref{eqn:equate-approx-exact-AGP} meaningful at small $\omega$, where any finite degree polynomial vanishes while $1/\omega$ diverges. One can consider other choices as well, see e.g. Ref.~[\onlinecite{lim2024definingclassicalquantumchaos}]. For example, if there is a hard ground state gap $\Delta$ another choice of this regularization might be to choose $\mu=\Delta$ and $g(x)=\theta(|x|-1)$, where $\theta(x)$ is the Heaviside step function. As it becomes clear from the following discussion our analysis does not depend on the details of $g(x)$. Clearly Eq.~\eqref{eqn:equate-approx-exact-AGP} also has to be regularized in the UV limit $\omega\to \infty$ where conversely any polynomial diverges, while $1/\omega\to 0$. This regularization will turn out to control the convergence of the AGP ansatz in Krylov space. 

In many-particle systems the spectrum is exponentially dense in the system size and hence can be regarded as continuous. In practice, modulo the subtleties related to taking the limits $\omega\to 0$ and $\omega\to\infty$, we see that the problem of finding the AGP in Krylov space maps to that of approximation of the function $1/\omega$ by odd polynomials in $\omega$. Thus, the variational method can be formulated as least squares minimization of the following \textit{cost function} 
\begin{equation} \label{eqn:cost-func}
    \mathcal{C}_\ell = \int d\omega \left(\sum_{k=1}^\ell \alpha_k \omega^{2k-1} + {1 \over \omega}\, g(\omega/\mu) \right)^2 \Phi_\lambda(\omega).
\end{equation}

\noindent where the squares of the matrix elements $\langle m \vert \partial_\lambda H \vert n \rangle$ set the relative weights for different frequencies via $\Phi_\lambda(\omega)$. Then, the coefficients $\alpha_k$ are found as
\begin{equation}
    \{\alpha_k\} = \min_{\alpha_k} \mathcal{C}_\ell
\end{equation}
We note that this cost function $\mathcal{C}_\ell$ is almost identical to that used in the standard variational approach, which minimizes the action~\eqref{eqn:variational_action}, up to an additional multiplication by $\omega$ of the expression in the brackets in that case. This minor point is discussed in greater detail in Appendix \ref{appendix:omega-cost-func}, but it does not significantly change the results.

Crucially, in Eq. \eqref{eqn:cost-func} all of the information about the particular system enters only via the spectral function $\Phi_\lambda(\omega)$. For a universal CD protocol we would like to develop a variational procedure which is independent of the details of the system. To do this, we will replace the exact spectral function with the following rectangular function:

\begin{equation} \label{eqn:phi-assumption}
    \Phi_{\rm step}(\omega)= 
    \begin{cases}
        0 & \text{for } 0 \leq \omega < \mu \\
        1 & \text{for } \mu \leq \omega \leq \Omega \\
        0 & \text{ for } \omega > \Omega
    \end{cases}
\end{equation}

Here $\mu$ is the low-frequency cutoff, which plays the same role as the cutoff in the filter function $g(\omega/\mu)$, so we keep the same notation. It determines the lowest frequency scale below which the approximate AGP does not follow the exact one. Likewise, the scale $\Omega$ plays the role of the upper frequency cutoff.

\begin{figure}[ht]
    \centering
    \includegraphics[width=\linewidth]{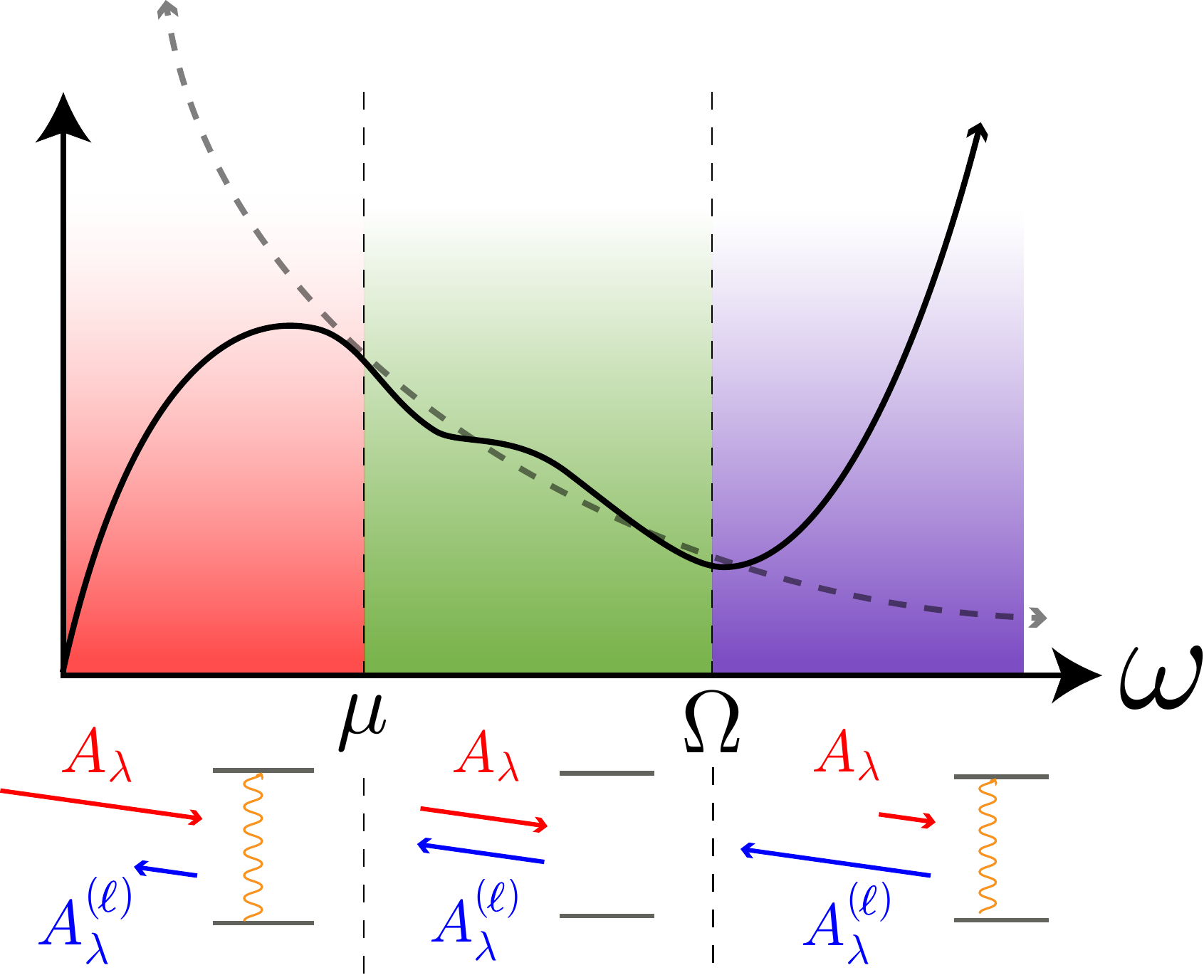}
    \caption{A cartoon illustration of the problem of fitting $1/\omega$ by odd polynomials. There are three regions: $\omega < \mu$ (red, IR), where the fit is poor and the approximate AGP $A_\lambda^{(\ell)}$ fails to replicate the true AGP $A_\lambda$, and so does not suppress low energy excitations. Next, $\mu \leq \omega \leq \Omega$ (green) where the approximation is good and so excitations are suppressed. Finally, $\omega > \Omega$ (violet, UV) where high energy excitations are excited, not by the finite rate of changing a parameter, but by the large error in the CD drive approximation of Eq. \eqref{eqn:commutator_ansatz}.}
    \label{fig:cartoon-fitting}
\end{figure}

The universal variational AGP construction will converge if by increasing the order of approximation $\ell$, we are able to decrease the low frequency cutoff $\mu$ (i.e. suppress increasingly low energy excitations) while keeping $\Omega$ sufficiently large such that the exact cost function~\eqref{eqn:cost-func} remains small. The advantage of the universal approach is that the information about the system only enters through the scaling of the exact spectral function in the limits $\omega\sim \mu\to 0$ and $\omega\sim \Omega\to \infty$, which in turn determine the scaling of the error with $\ell$. The variational optimization becomes a purely mathematical problem of fitting the function $-1/\omega$ by odd polynomials in $\omega$, within the range $\mu \leq \omega \leq \Omega$. This procedure is schematically illustrated in Figure \ref{fig:cartoon-fitting}. The local CD protocol fails to suppress transitions to the states with $|\omega_{mn}|<\mu$ and conversely \textit{induces} transitions at $|\omega_{mn}|>\Omega$. After the optimal fitting in this window is achieved by e.g. least squares the parameters $\mu$ and $\Omega$ have to be optimized for each $\ell$. In this way the number of independent variational parameters is reduced from $\ell$ to only two.

The expansion for the AGP in Eq. \eqref{eqn:commutator_ansatz} in terms of nested commutators leads to the polynomial approximation of the function $1/\omega$ from Eq.~\eqref{eqn:cost-func}. It is more convenient instead to use an expansion in different polynomials. The reason is that the monomials of Eq. \eqref{eqn:approx-AGP-mtx-elts} are difficult to work with numerically due to the exponential growth of the coefficients $\alpha_k$. In the direct variational approach it is  most convenient to work with orthogonal operators forming a Krylov space~\cite{selsThermalizationDiluteImpurities2023,bhattacharjeeLanczosApproachAdiabatic2023, takahashiShortcutsAdiabaticityKrylov2024}. It is straightforward to check that this procedure is equivalent to an approximation of $1/\omega$ in Eq.~\eqref{eqn:cost-func} in terms of polynomials which are orthogonal with respect to the spectral function $\Phi_\lambda(\omega)$, which depends on the details of the system. Thus it is not suitable for a universal protocol. Instead we replace Eq. \eqref{eqn:commutator_ansatz} with a linear combination of model independent orthogonal polynomials. In this work we consider an expansion in terms of Chebyshev polynomials $T_n(\omega/\Omega)$. We rescale the frequency $\omega$ because the Chebyshev polynomials $T_n(x)$ are well-behaved and orthogonal in the interval $x\in[-1,1]$ with the weight function $1/\sqrt{1-x^2}$~\cite{Trefethen2019}. In the operator language this expansion corresponds to replacing the ansatz~\eqref{eqn:commutator_ansatz} with an equivalent one:

\begin{equation}
  \label{eqn:commutator_ansatz_tchebyshev}
    A_\lambda^{(\ell)} = i \sum_{k = 1}^\ell \beta_k T_{2k-1}(\hat {\mathcal L} / \Omega)\,\partial_\lambda H / \Omega
\end{equation}
where $\hat {\mathcal L}$ stands for the Liouvillian super-operator: $\hat {\mathcal L} O=[H,O]$. It is clear that the set of coefficients $\beta_k$ and $\alpha_k$ are uniquely related for any $\ell$. More detail about this construction is given in Appendix \ref{appendix:chebyshev-AGP}.

For the universal ansatz we optimize the cost function~\eqref{eqn:cost-func}, where we replace the true spectral function $\Phi_\lambda(\omega)$ with the rectangular one~\eqref{eqn:phi-assumption}. This means that we have cast the problem as choosing the coefficients $\beta_k$ which minimize the following cost function
\begin{equation} \label{eqn:cost-fn-chebyshev}
    \tilde{\mathcal C}_\ell(\zeta) = \int_{\zeta}^1 dx \left(\sum_{k=1}^\ell \beta_k T_{2k-1} (x) + {1 \over x} \right)^2,
\end{equation}

\noindent where $x$ plays the role of $\omega/\Omega$ and $\zeta=\mu/\Omega$. Here we have also replaced $g(x) \rightarrow 1$ since the regularization is now provided by the lower cutoff $\zeta$. For fixed values of $\ell$ and $\zeta$ we can now minimize Eq.~\eqref{eqn:cost-fn-chebyshev} with respect to the coefficients $\beta_k$ and plug them into~\eqref{eqn:commutator_ansatz_tchebyshev} to find the local ansatz for the AGP. For each $\ell$, this depends on two parameters: $\mu=\zeta \Omega$ and $\Omega$.
Then, all that remains is to find an optimal way to choose them.

An accurate polynomial approximation of $1/\omega$ in the interval $[\mu,\Omega]$ ensures that the main contributions to Eq.~\eqref{eqn:cost-func} comes from the regions $\omega\lesssim \mu$ and $\omega>\Omega$. The first, low frequency contribution is determined by the minimum gap, and more generally by the density of the low-energy states. If optimal value of $\mu(\ell)$ is a monotonically decreasing function of $\ell$, the low frequency contribution is suppressed more and more as $\ell\to\infty$. The latter, high frequency contribution is clearly determined by the high frequency tail of $\Phi_\lambda(\omega)$, which, in the thermodynamic limit, is typically given by
\begin{equation} \label{eqn:spec-fn-decay}
    \Phi_\lambda(\omega) \sim \exp[-\Gamma \vert \omega\vert^{\alpha}]
\end{equation}
It has been rigorously proven for systems with a bounded local Hilbert space such as spins or lattice fermions, which satisfy Lieb-Robinson bounds, that $\alpha \geq 1$\cite{abaninExponentiallySlowHeating2015}. It was further conjectured that for generic systems, finite temperature spectral functions saturate the bound $\alpha = 1$~\cite{parkerUniversalOperatorGrowth2019}, with an additional logarithmic correction in one-dimensional systems~\cite{ avdoshkinEuclideanOperatorGrowth2020}. Similar exponential scaling of $\Phi_\lambda (\omega)$ was numerically observed for generic classical systems (both integrable and chaotic) with an unbounded spectrum~\cite{lim2024definingclassicalquantumchaos}.

The remainder of this work will be concerned with how the two parameters $\mu(\ell)$ and $\Omega(\ell)$ for the optimal protocol scale with the order of approximation $\ell$. As we shall see, not only $\Omega(\ell)$, but surprisingly also $\mu(\ell)$ depends strongly on the exponent $\alpha$: for $\alpha < 1$ the universal protocol does not exist in the thermodynamic limit in the sense that the UV error introduced by using higher order polynomials is too big to allow for $\mu(\ell)$ to decrease with $\ell$. Conversely for $\alpha>1$ the universal protocol works such that $\mu(\ell)\sim \ell^{1/\alpha-1}$ decreases with $\ell$. The generic borderline case $\alpha=1$ requires extra care and one can achieve very slow convergence  $\mu(\ell) \sim 1/\log(\ell)$~\cite{finžgar2025counterdiabaticdrivingperformanceguarantees}. In the remainder of this paper we will explain where this result comes from and illustrate the emerging scaling of $\mu(\ell)$ and $\Omega(\ell)$ for three different situations.

\subsection{Noninteracting models} \label{subsec:fit-range}

Let us first consider the case of a many-body system which may be reduced to a system of free particles with a bounded single-particle spectrum. These models, for example, are naturally realized in lattice systems within the tight-binding approximation. 
In this case, there is a maximum excitation energy $\Omega_{\rm max}$ beyond which all matrix elements of local single-particle operators are identically equal to zero: $\Phi(\omega) = 0$ for $\omega > \Omega_{\rm max}$. It is then clear that we can set the upper cutoff $\Omega=\Omega_{\rm max}$ irrespective of $\ell$ and never worry about accuracy of the polynomial fit of $1/\omega$ above this scale. Because $\Omega$ is independent of $\ell$ we only need to optimize for $\mu(\ell)$, and hence find the optimal $\zeta(\ell)=\mu(\ell)/\Omega$.

As a specific example, we choose the transverse field Ising (TFI) model, which is exactly solvable by mapping to free fermions via the Jordan-Wigner transformation \cite{jordanUeberPaulischeAequivalenzverbot1928}. The protocol is to anneal from the ground state in the paramagnetic phase to the ferromagnetic phase. This is encoded in the following Hamiltonian

\begin{equation} \label{eqn:tfim-anneal}
    H(\lambda) = \lambda (-J \sum_i \sigma_i^z \sigma_{i+1}^z ) + (1-\lambda) (-h\sum_i \sigma_i^x)
\end{equation}

\noindent where tuning $\lambda$ from 0 to 1 induces this phase transition. In this model, $\Omega_{\rm max} = {\rm max}_{\lambda} \left[4(\lambda J + (1-\lambda) h)\right]$, and we take $J = h = 1$ so that $\Omega = \Omega_{\rm max} = 4$. 

We can choose the low frequency cutoff $\mu$ so that the final state fidelity of the universal protocol
\begin{equation} \label{eqn:final-fid}
    \mathcal{F} = \vert \langle \psi(t = t_{final}) \vert \psi_0(t=t_{final}) \rangle \vert^2
\end{equation}

\begin{figure}
    \centering
    \includegraphics[width=\linewidth]{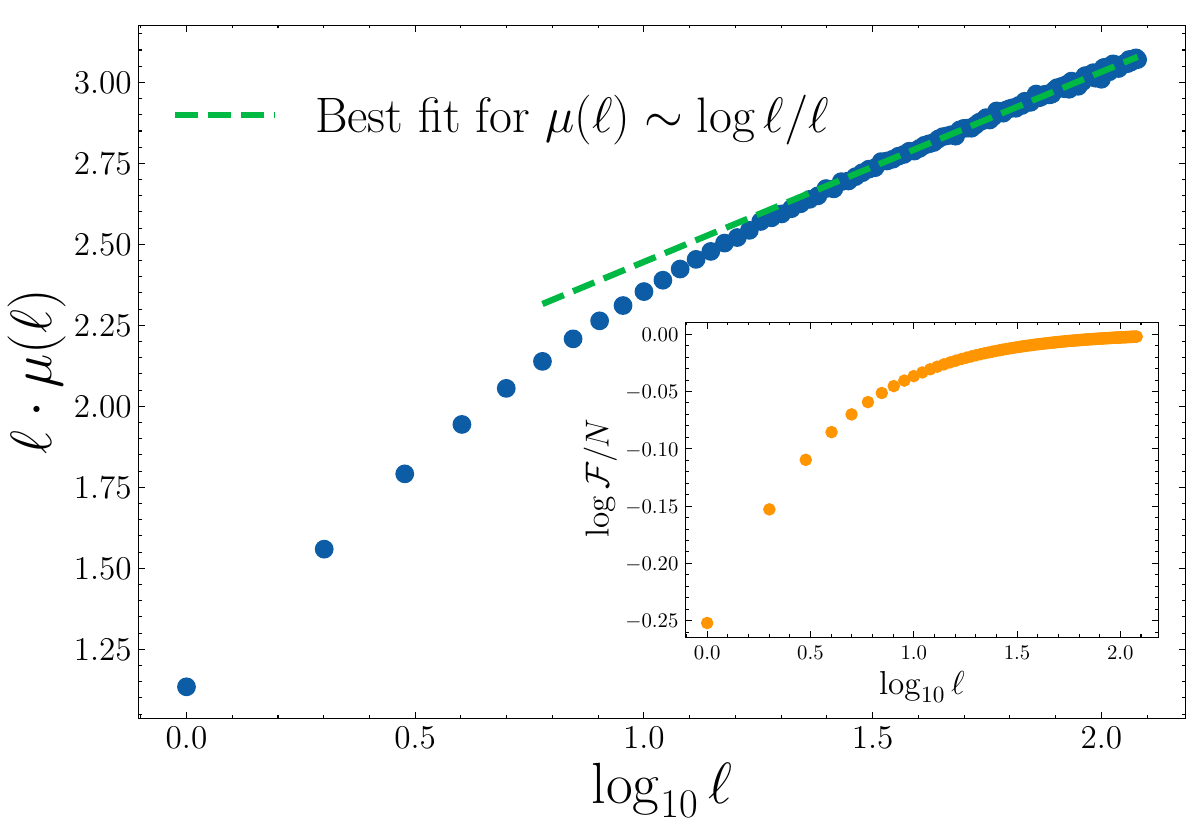}
    \caption{For the transverse field Ising model, with $N = 500$, we compute the $\mu(\ell)$ which maximizes fidelity~\eqref{eqn:final-fid}. We observe the asymptotic scaling of $\mu(\ell) \sim (\log \ell) / \ell$, which is consistent with known mathematical results for the polynomial fitting of $1/\omega$~\cite{childsQuantumAlgorithmSystems2017}. As $\mu(\ell) \rightarrow 0$, low energy excitations are suppressed and we approach unit fidelity. This is seen in the inset when the scaled fidelity $\log\mathcal{F}/N \rightarrow 0$ as $\ell\to\infty$.}
    \label{fig:lower-window-conv}
\end{figure}
\begin{figure}
    \centering
    \includegraphics[width=\linewidth]{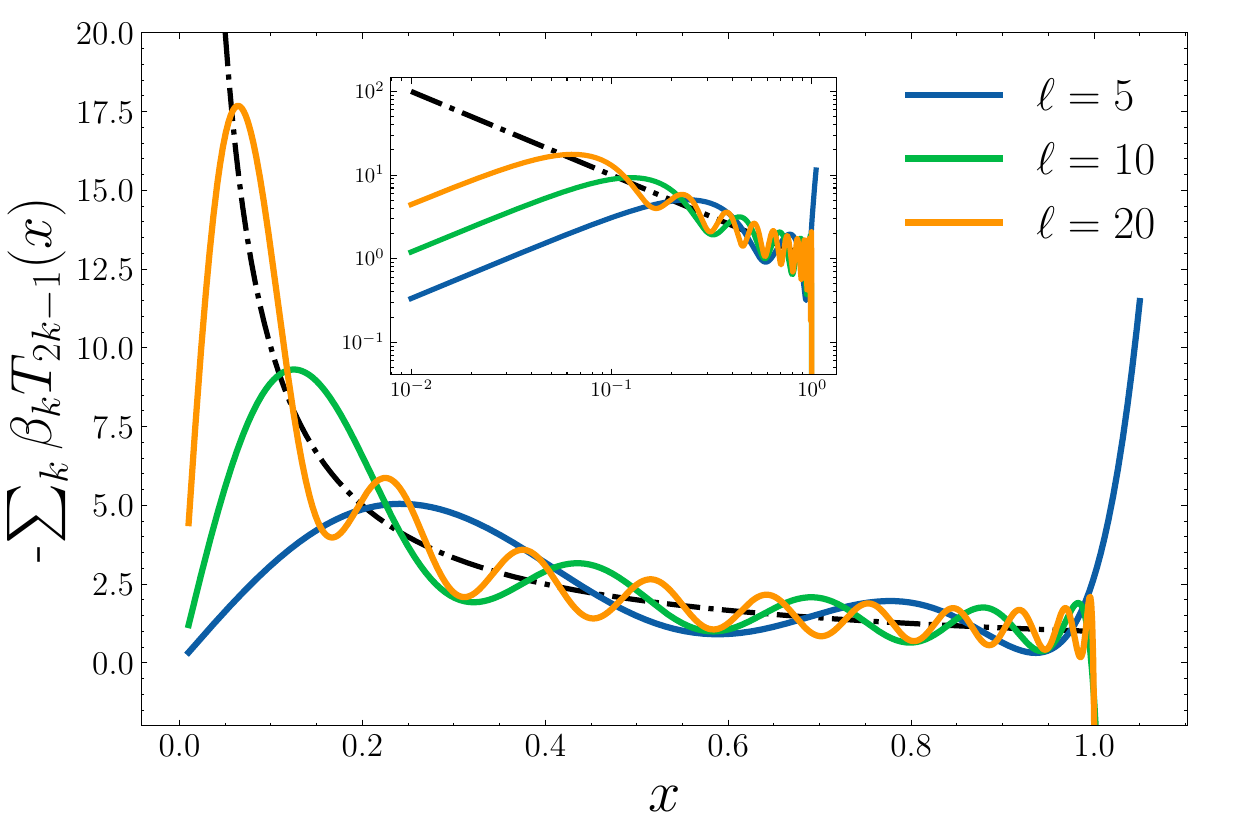}
    \caption{Approximation of  $1/x$ by Chebyshev polynomials, where $x = \omega/\Omega$. The inset shows the same plot on a log scale. In each case, the optimal $\mu(\ell)$ that is obtained by maximizing fidelity for the TFI model annealing problem is used, with fixed $\Omega(\ell) = \Omega_{max}$.}
    \label{fig:1/x-fit}
\end{figure}


\noindent is maximized, where $\vert \psi(t = t_{final}) \rangle$ is the state evolved by $H_{CD}$ in Eq. \eqref{eqn:H_CD} with the approximate AGP obtained by the polynomial fitting method, and $\vert\psi_0(t=t_{final})\rangle$ is the ground state of $H(\lambda = 1)$. Because this model consists of free particles, the fidelity may be calculated directly in the thermodynamic limit where finite size effects are not a concern. We then numerically optimize this fidelity with respect to $\mu(\ell)$ for a fixed $\Omega=4$, which in turn determines the variational coefficients $\beta_k(\ell)$ as they depend implicitly on $\mu(\ell)$ through the cost function. We then find the asymptotic scaling $\mu(\ell) \sim (\log(\ell) / \ell) \Omega$, shown in Figure \ref{fig:lower-window-conv}. The resulting fits to $1/x$ for a few values of $\ell$ are shown in Fig. \ref{fig:1/x-fit}. Instead of maximizing the fidelity one can use rigorous mathematical results for polynomial approximation of $1/x$ using the cost function $\tilde {\mathcal  C}_\ell$ in Eq.~\eqref{eqn:cost-fn-chebyshev} within a finite window \cite{childsQuantumAlgorithmSystems2017} and arrive at the same asymptotic result.

Once the optimal value of the ratio 
\begin{equation} \label{eqn:ratio-scaling}
    \zeta(\ell)\equiv {\mu(\ell) \over \Omega(\ell)} \sim C {\log\ell \over  \ell}
\end{equation}
together with the coefficients $\beta_k$, are fixed by using the test TFI model, the only remaining scale to be optimized for general models is the UV cutoff $\Omega(\ell)$. In Sec.~\ref{sec:examples} we will do this optimization numerically, again using the fidelity~\eqref{eqn:final-fid}. However, as we discuss next, one can understand the asymptotic behavior of $\Omega(\ell)$ (and hence of $\mu(\ell)$) analytically using only the high frequency scaling of the spectral function~\eqref{eqn:spec-fn-decay}. We highlight again that the TFI model here serves merely as a means of finding a universal protocol up to an overall scale $\Omega(\ell)$.

\subsection{Interacting Models \& Asymptotes of the Spectral Function} \label{subsec:asymptotes}

 For single-particle models, where there are no excitations above a certain maximum, this method is guaranteed to converge since for constant $\Omega(\ell)$, clearly $\lim_{\ell\to\infty} \mu(\ell)= 0$. Thus, the interval $[\mu(\ell), \Omega]\xrightarrow[]{\ell \to\infty}[0,\Omega]$ such that the universal protocol will asymptotically suppress more and more low-energy excitations with increasing $\ell$ without exciting any high-frequency modes. In general, the convergence of the universal protocol with $\ell$ is guaranteed as long as $\Omega(\ell)$ grows slower than $\ell/\log(\ell)$. In this case we still have $\lim_{\ell\to\infty} \mu(\ell)= 0$.

For a generic interacting model we can estimate the scaling $\Omega(\ell)$ from the requirement that the UV part of the cost function does not increase with $\ell$~\eqref{eqn:cost-func}. At high frequencies this part is dominated by the highest order polynomial $k=2\ell-1\approx 2\ell$ so we can estimate 
\begin{equation} \label{eqn:cost-func-asymp}
    \mathcal{C} \sim \alpha_\ell^2 \Omega^{4\ell} \exp[- \Gamma \vert\Omega\vert^{\alpha}],
\end{equation}

where we assumed that Eq.~\eqref{eqn:spec-fn-decay} holds at sufficiently large frequencies. The condition that $\mathcal{C} \ll 1$ thus requires that 

\begin{equation}
\label{eqn:Omega_ell}
    \Omega\gg \left({4\ell\over \Gamma} \log \left(4\ell\over \Gamma\right)\right)^{1/\alpha}\sim (\ell\log \ell)^{1/\alpha}
\end{equation}
Note that in one-dimensional models, asymptotically the spectral function has an extra logarithmic correction~\cite{avdoshkinEuclideanOperatorGrowth2020}: $\Phi(\omega)\sim \exp[-\Gamma \omega\log(\Gamma \omega)]$. It is easy to see that this correction modifies Eq.~\eqref{eqn:Omega_ell} to $\Omega\gg \ell / \Gamma$.

We see that in order to control the UV error of the cost function we should choose 
$\Omega(\ell)$ to be at least as large as the bound given by Eq.~\eqref{eqn:Omega_ell}. As a result, using Eq.~\eqref{eqn:ratio-scaling} and ignoring for simplicity logarithmic corrections we arrive at the scaling
\begin{equation}
\label{eqn:mu_ell_scaling}
    \mu(\ell)\sim \ell^{{1\over \alpha}-1}.
\end{equation}
Physically, this result means that for $\alpha\leq 1$ the error in approximating the AGP, which results in high energy excitations, is very large unless $\Omega(\ell)$ increases very quickly with $\ell$. In turn, this requisite rapid increase while keeping a fixed ratio $\mu(\ell)/\Omega(\ell)$ does not allow decreasing $\mu(\ell)$, so that the local protocol will not suppress low-energy, non-adiabatic transitions at large $\ell$. Conversely for $\alpha>1$ the necessary increase of $\Omega(\ell)$ to control the UV error is sufficiently slow such that $\mu(\ell)$ decreases and the local CD protocol works better and better with increasing $\ell$.

Summarizing our discussion here we arrive at the most surprising conclusion in this work, tying the complexity of local CD protocols to the high frequency properties of the system. This happens despite the fact that when constructing counterdiabatic protocols, we are typically interested in suppressing the low energy excitations in the system (which set the norm of the exact AGP in Eq. \eqref{eq:chi_lambda}). The same low energy excitations determine the minimal protocol time in conventional quantum annealing schemes. For example, if we analyze driving across a critical point, such low-energy excitations get ``frozen out'' in the Kibble-Zurek mechanism \cite{zurekCosmologicalExperimentsSuperfluid1985,Polkovnikov_2005,Dziarmaga_2005,Zurek_2005}. The scaling~\eqref{eqn:mu_ell_scaling} states that contrary to these expectations, performance of the local CD protocol -- at least in the Krylov space -- is tied to the high frequency response of the system. This suggests there is some fundamental connection between the short- and long-time dynamics as alluded to earlier. This point will be further developed in Section \ref{sec:long-short-time-connection}.

Note that our results are asymptotic in nature and do not imply that approximate CD driving cannot lead to significant improvements in fidelity at finite orders $\ell$. There are many examples in the literature showing that this is the case in various interacting models, both quantum and classical. Also, for finite systems the spectral function $\Phi_\lambda(\omega)$ generally saturates at $\omega\propto N$, where $N$ is the system size, suggesting that even for the generic case $\alpha=1$ the performance of local CD protocols will inevitably start improving for $\ell>\ell^\ast\sim N$.

\section{Constructing Universal Protocols for Specific Systems} \label{sec:examples}

We will now discuss three separate models, each with different high-frequency behavior of the spectral function. i) The noninteracting transverse field Ising model with disorder, which has a high frequency hard cutoff; ii) a generic interacting model with approximately exponential  ($\alpha = 1$) high frequency tail of the spectral function (modulo log-correction) and; iii) an interacting integrable model which has an approximately Gaussian high frequency tail corresponding to $\alpha=2$.

\subsection{Transverse Field Ising Model with Modulated Fields} \label{subsec:TFIM-disorder}

To test our universal protocol, we introduce disorder in the fields $h_i$ of the TFI model, which retains the paramagnetic to ferromagnetic transition present in the clean model \cite{fisherRandomTransverseField1992}. Instead of analyzing a fully disordered model, we split the spin chain into contiguous blocks $\mathcal{B}$, each of size $N_{\mathcal{B}} = 4$. There is random disorder within each block, which is repeated over the lattice. The Hamiltonian is

\begin{equation} \label{eqn:tfim-anneal}
    H(\lambda) = \lambda (-J \sum_i \sigma_i^z \sigma_{i+1}^z ) + (1-\lambda) (-\sum_{\mathcal{B}} \sum_{i\in\mathcal{B}} h_i \sigma_i^x)
\end{equation}

\noindent with $h_i$ drawn uniformly between $0$ and $h$ with $h = J = 1$. This will increase the size of the blocks in momentum space after the Jordan-Wigner transformation, but the blocks remain small enough that the dynamics may be computed via exact diagonalization.

The presence of randomness does not affect the maximum single particle energy, which remains $\Omega_{max} \leq \max_\lambda 4(\lambda J + (1-\lambda)h)$ so we again choose $\Omega = \Omega_{max} = 4$.

\begin{figure}
    \centering
    \includegraphics[width=\linewidth]{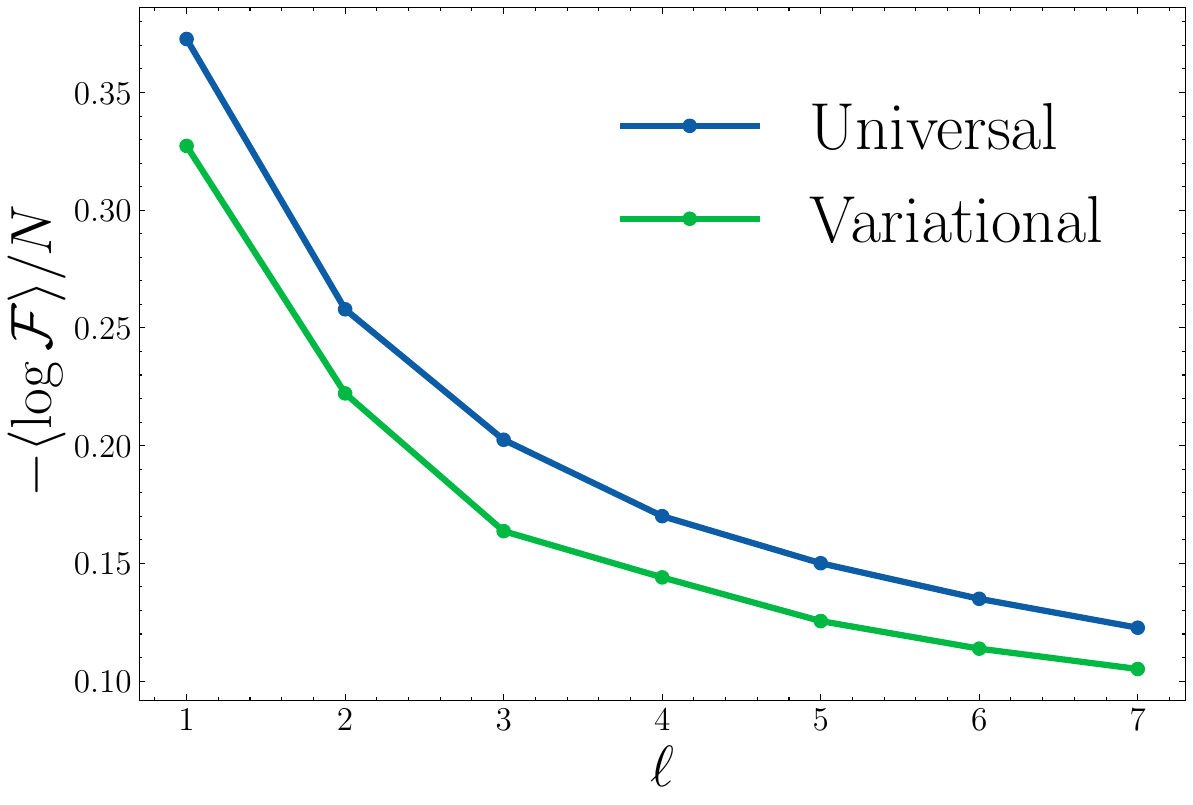}
    \caption{ The fidelity density of the final state in the thermodynamic limit after annealing across the Ising transition in the presence of modulated fields. While the variational approach is slightly better due to precise knowledge of the spectral function, the universal protocol offers competitive performance using only knowledge of the local energy scale. A similar plot for a non-integrable model at finite system size in shown in Fig. \ref{fig:compare-cost-funcs} in Appendix \ref{appendix:omega-cost-func}.}
    \label{fig:tfim-disord}
\end{figure}

Using $\zeta(\ell)=\mu(\ell)/\Omega(\ell)$ found for the non-disordered TFI model, with the large $\ell$ asymptote given by Eq. \eqref{eqn:ratio-scaling}, we show in Figure \ref{fig:tfim-disord} the results of the protocol obtained for the clean model applied to this model. Despite having no knowledge of the specific details of the disorder present in the system, the fidelity density in the thermodynamic limit improves at a rate similar to the variational approach.

\subsection{Ising Model with Next-Nearest-Neighbor Interactions} \label{subsec:NNN-Ising}

Next we move to analysis of a generic nonintegrable system by adding next-nearest-neighbor interactions to the (clean) TFI model. This corresponds to driving the following Hamiltonian:

\begin{multline} \label{eqn:NNN-tfim}
    H(\lambda) = \lambda (-J \sum_i \sigma_i^z \sigma_{i+1}^z -J_2 \sum_i \sigma_i^z \sigma_{i+2}^z ) \\
    + (1-\lambda) (-h\sum_i \sigma_i^x)
\end{multline}

\noindent where we take $J_2 / J = 0.25$, which will retain the same transition, albeit at a different value of $\lambda_c$ \cite{suzukiQuantumIsingPhases2013}.

\begin{figure}
    \centering
    \includegraphics[width=\linewidth]{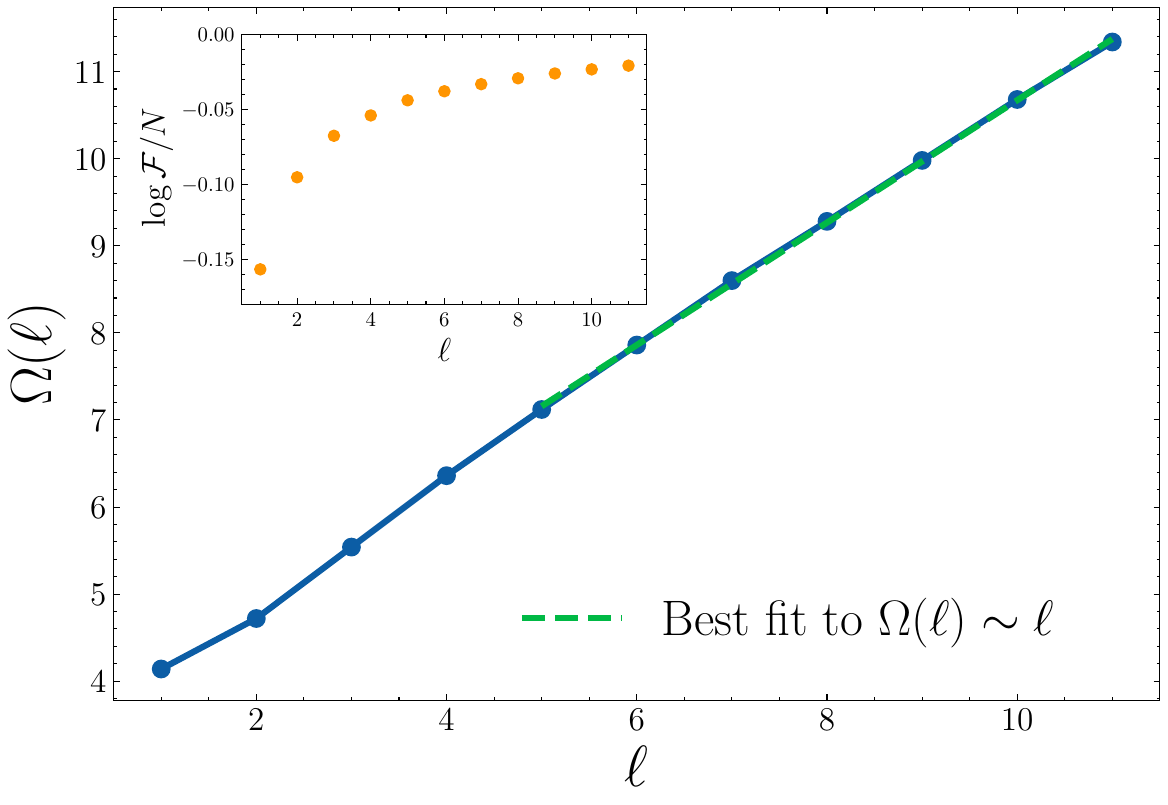}
    \caption{For the next-nearest-neighbor Ising model with $N = 14$, annealing from the paramagnetic to ferromagnetic phases, the optimal $\Omega(\ell)$ is obtained numerically by maximizing the fidelity defined in Eq. \eqref{eqn:final-fid}. The dashed line is a linear fit, which is consistent with a general argument from Sec.~\ref{subsec:asymptotes}. The inset shows the scaled fidelity of the final state. Although the asymptotic behavior of the fidelity cannot be reliably obtained due to finite-size effects, it is expected to plateau at a small non-zero value in the thermodynamic limit $N\to \infty$.}
    \label{fig:NNN-tfim-window}
\end{figure}

From our discussion in Section~\ref{subsec:asymptotes}, for generic one-dimensional systems, one expects an asymptotic linear scaling of the UV cutoff: $\Omega(\ell) \sim \ell$. In Figure \ref{fig:NNN-tfim-window} we show the scaling of $\Omega(\ell)$ obtained by choosing it to maximize the final state fidelity at fixed $\zeta(\ell)=\mu(\ell)/\Omega(\ell)$. We find very good agreement with that prediction.

With the scaling of $\zeta(\ell)$ from Eq. \eqref{eqn:ratio-scaling}, this linear growth of $\Omega(\ell)$ gives $\mu(\ell) \sim \log\ell$. Therefore the infrared cutoff, instead of decreasing, slowly increases with $\ell$. We therefore conclude that in the thermodynamic limit $N\to\infty$ the universal protocol, and more generally the Krylov space variational ansatz for the AGP does not converge. Additionally, the universal protocol still leads to a significant improvement of fidelity at finite $\ell$.

\subsection{XXZ Model} \label{subsec:XXZ}

It was conjectured that certain non-generic interacting integrable models can have slower operator growth than generic ones, and hence have a faster-than-exponential decay of the spectral function~\cite{parkerUniversalOperatorGrowth2019}. In particular, there is numerical evidence of Gaussian decay, corresponding to $\alpha = 2$ in Eq. \eqref{eqn:spec-fn-decay}, for the XXZ model \cite{elsayedSignaturesChaosTime2014, leblondEntanglementMatrixElements2019}. In Appendix \ref{appendix:xxz-gaussian-decay}, we provide further numerical evidence confirming this decay by looking at the growth of Lanczos coefficients. To test the scaling of $\Omega(\ell)$ in this case we now analyze this model using the following annealing Hamiltonian:

\begin{equation} \label{eqn:xxz_anneal}
    H(\lambda) = (1-\lambda) (-J/3 \sum_i \vec{\sigma}_i \cdot \vec{\sigma}_{i+1}) + \lambda (-\Delta \sum_i \sigma_i^z \sigma_{i+1}^z)
\end{equation}
where $\vec{\sigma} = (\sigma^x, \sigma^y, \sigma^z)$. We take $J = \Delta = 1$, so that we are annealing from the Heisenberg point to a pure Ising point. 

\begin{figure}
    \centering
    \includegraphics[width=\linewidth]{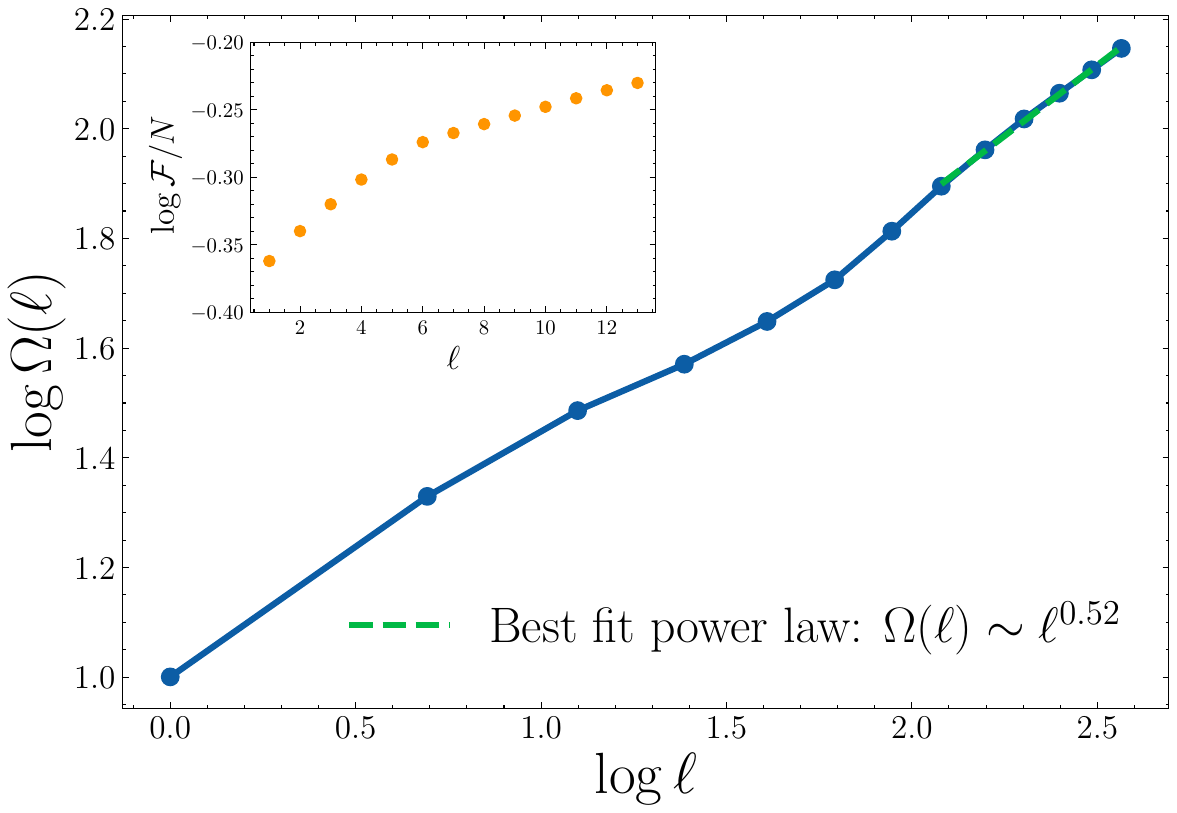}
    \caption{For the XXZ model with $N = 18$, annealing from the Heisenberg to Ising point, the optimal $\Omega(\ell)$ is obtained numerically by maximizing the fidelity as before. The dashed line is the best fit to a power-law of the data, showing a slope 0.52 which is very close to the  Section \ref{subsec:asymptotes} prediction of 0.5 with a logarithmic correction. The asymptotic behavior of the fidelity, inset, is hard to predict for the finite sized systems accessible. But it does not appear to plateau.}
    \label{fig:XXZ-window}
\end{figure}

The scaling argument from Sec.~\ref{subsec:asymptotes} for $\alpha = 2$ gives $\Omega(\ell) \sim \sqrt{\ell}$ up to log-corrections. In Figure \ref{fig:XXZ-window}, we show that the same approach of choosing $\Omega(\ell)$ with fixed $\zeta(\ell)$ to maximize the final state fidelity indeed quite accurately reproduces this scaling. In turn, from Eq.~\eqref{eqn:mu_ell_scaling} we see that $\mu(\ell) \sim 1/ \sqrt{\ell}$, suggesting that the fidelity of the universal protocol should be improving with increasing $\ell$ as is evident from the inset in Fig.~\ref{fig:XXZ-window}.

\section{Extracting hydrodynamic tails from Lanczos coefficients} \label{sec:long-short-time-connection}

Across a wide variety of interacting systems, the long-time dynamics may often be described in some universal way, such as e.g. charge or energy diffusion in systems with corresponding conservation laws~(see Ref.~\onlinecite{RevModPhys.49.435} for a review). Recently, there has been extensive research on so-called generalized hydrodynamics where analogous universal long-time dynamics are seen in integrable models.~\cite{doyonLectureNotesGeneralised2020, bertiniTransportOutofEquilibriumChains2016, malvaniaGeneralizedHydrodynamicsStrongly2021}

Another area of active investigation is operator spreading. In particular, in a series of works it was argued that short-time dynamics of a quantum system in Krylov space is universal and can be characterized by the Krylov complexity~\cite{parkerUniversalOperatorGrowth2019, avdoshkinEuclideanOperatorGrowth2020, rabinoviciOperatorComplexityJourney2021}. In this approach time evolution of an observable in quantum (or classical) Hamiltonian system can be thought of as a one-dimensional hopping problem in the Krylov operator space formed from nested commutators of the operator describing this observable and the Hamiltonian like those entering Eq.~\eqref{eqn:commutator_ansatz}. The hopping amplitudes are then given by the so called Lanczos coefficients appear in the Gram-Schmidt orthonormalization of these commutators (see e.g. Ref.~\onlinecite{loizeau2025openingkrylovspaceaccess} for details). Formally all Lanczos coefficients can be found by analyzing a short time expansion of local operators in the Heisenberg representation. Finding if there are any connections between this short time evolution and long time hydrodynamic tails is an open and long-standing  problem. 

Unexpectedly, our work provides an insight into this. In particular, for systems whose UV spectral function tails decay faster than exponentially (corresponding to sublinear growth of Lanczos coefficients) using the universal AGP we can reconstruct the hydrodynamic tails of the spectral function, and hence the long-time asymptotic behavior of the correlation functions.  Conversely, in systems with linear Lanczos coefficient growth, corresponding to exponential tails of the spectral function, the hydrodynamic tails could be extracted from the universal AGP only when the number of the Lanczos coefficients exceeds the system size. While we do not claim that there is no other method which can reveal such long time information from the Lanczos coefficients, it seems that the UV tails of the spectral function play a similar fundamental role in accessing this low frequency information as they do in constructing the universal CD protocol.

Now let us briefly outline how one can extract the hydrodynamic tails from the universal CD protocol. If we regard $\partial_\lambda H$ as a perturbation arising from changing $\lambda$, then the commutators $\hat{\mathcal{L}}^{2k-1} \partial_\lambda H$ which appear in the approximate AGP simply correspond to a short-time expansion of the evolution of this perturbation in the interaction picture. We can combine these with the coefficients $\alpha_k$ or $\beta_k$, which are agnostic to any physics, to compute a universal AGP. In this way we can determine the AGP norm (or fidelity susceptibility) using only these universal coefficients and the nested commutator norms (which are uniquely related to the Lanczos coefficients). In turn from Eq.~\eqref{eq:chi_lambda} we see that the scaling of the fidelity susceptibility with $\mu$ is directly related to the low-frequency scaling of the spectral function (see also Refs.~\onlinecite{kimIntegrabilityAttractorAdiabatic2024, pandeyAdiabaticEigenstateDeformations2020}). This scheme thus works as long as by increasing the number of terms in the universal protocol we can effectively access lower values of $\mu$. The central result of our work is that $\mu(\ell)$ decreases in the thermodynamic limit for systems with high frequency tails decaying faster than exponentially.

Let us now illustrate these considerations using the the example of the XXZ chain discussed in Section \ref{subsec:XXZ} of the main text. There, since $\mu \sim 1/\sqrt{\ell}$ we expect that as $\ell$ is increased, the universal AGP should approach the true AGP. If the dynamics are diffusive as expected for the XXZ chain with $\Delta>1$~\cite{Gopalakrishnan_2019}, then $\Phi_\lambda(\omega)\sim 1/\sqrt{\omega}$~\cite{selsThermalizationDiluteImpurities2023} and for large enough $\ell$ according Eq. \eqref{eq:chi_lambda}) $\chi_\lambda$ should scale with $\mu^{-3/2}$. For the universal AGP, we can compute this directly from the definition in Eq. \eqref{eqn:approx-AGP-mtx-elts}:

\begin{align} \label{eq:approx-agp-norm}
    \Vert A_\lambda^{(\ell)} \Vert^2 & = \sum_{m, n} \left( \sum_k \alpha_k (\mu, \Omega) \ \omega_{mn}^{2k-1} \right)^2 \vert \langle m \vert \partial_\lambda H \vert n \rangle \vert^2 \\
     &= \int_{-\infty}^\infty d\omega \left( \alpha_k(\mu, \Omega) \  \omega^{2k-1} \right)^2 \Phi_\lambda(\omega)
\end{align}

\begin{figure*}
    \centering
    \includegraphics[width=\linewidth]{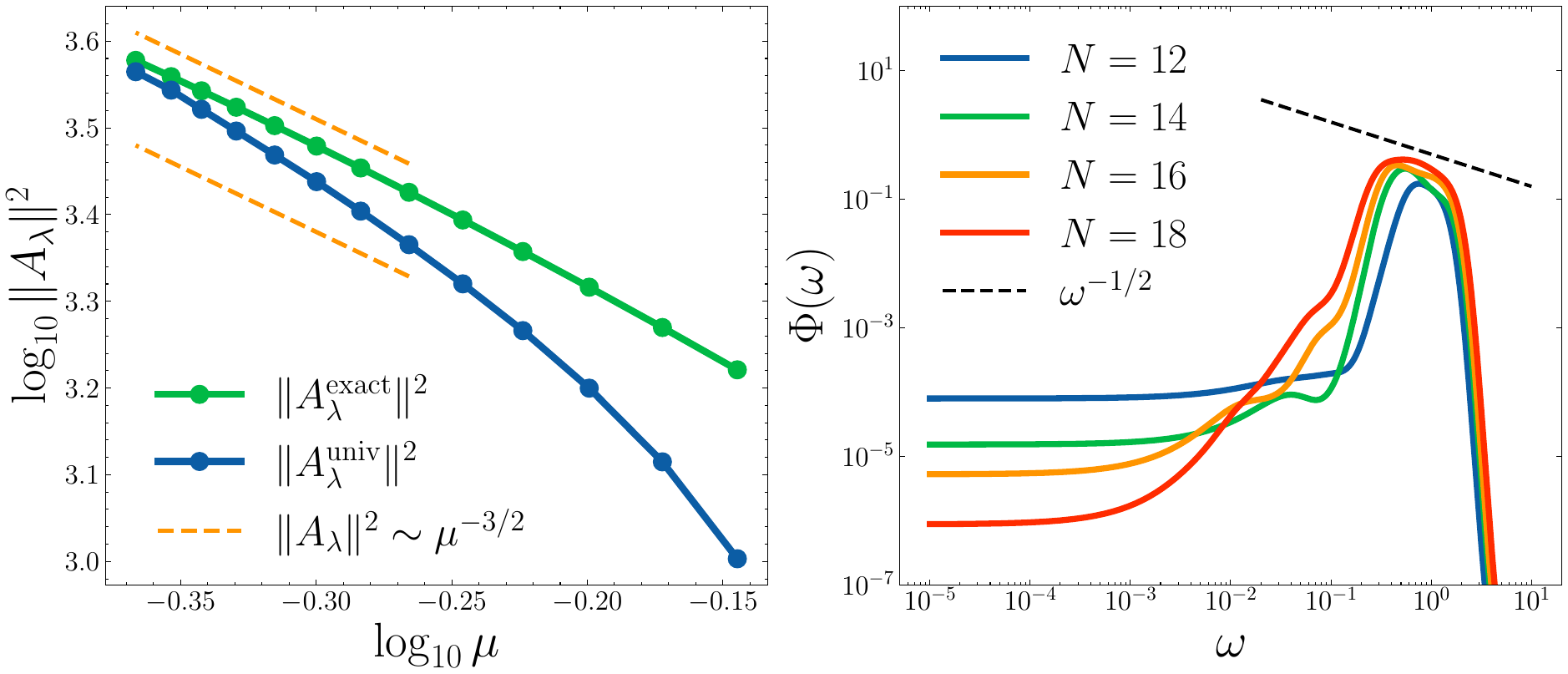}
    \caption{On the left hand side, the exact AGP norm, computed via Eq. \eqref{eq:chi_lambda} is compared with that from the universal AGP in Eq. \eqref{eq:approx-agp-norm}. The dashed lines show the expected asymptotic scaling for diffusion. The numerics are for the $N = 18$ XXZ chain at $\lambda = 3/8$, corresponding to $\Delta = 1.2$ in standard parametrization of XXZ model. Minor finite size effects may play a role below $\log_{10} \mu \approx -0.3$ since the largest operators in the commutator expansion reach the boundary modifying the commutator norms compared to thermodynamic limit. On the right hand side, the exact spectral function for this model, averaged over several nearby values of $\lambda$, is plotted for different system sizes. The regime of $\omega$ at which diffusion is present slowly increases with system size and is expected to extend all the way to zero frequency in the thermodynamic limit. For $\mu$ within this region, we expect the scaling $\Vert A_\lambda \Vert^2 \sim \mu^{-3/2}$.}
    \label{fig:norm-scaling}
\end{figure*}

Here, we have written $\mu$ and $\Omega$ as arguments of $\alpha_k$ explicitly to highlight that they alone determine the coefficients. Then, we can compare numerically the norms of the approximate and exact AGPs for each $\mu$. This is shown in Figure \ref{fig:norm-scaling}. From this, it appears that the universal AGP is able to capture the diffusive hydrodynamics of the model, despite the only input about the model being information present in a short time expansion. We expect that the scaling will work even better if instead of exact diagonalization one uses other numerical tools allowing access to short time dynamics (and hence the Lanczos coefficients) in larger systems.

Finally let us point that in the generic models (say a non-integrable Ising model discussed here) increasing the number of terms $\ell$ in the universal protocol does not lead to decrease of $\mu(\ell)$ as long as $\ell\ll L$.~\footnote{We mention again that in the parallel paper~\cite{finžgar2025counterdiabaticdrivingperformanceguarantees} it was argued that if one uses the information about high frequency tail of the spectral function one can make $\mu~\sim 1/\log(\ell)$ such that it slowly decreases with $\ell$.} This implies that we cannot access the low $\mu$ asymptote of $\chi_\lambda$ and hence of the spectral function unless $\ell$ exceeds system size $L$ and the upper cutoff $\Omega$ stops growing with $\ell$. 

\section{Summary \& Outlook} \label{sec:conclusions}

Exact CD driving allows for adiabatic state preparation at arbitrarily fast speeds. By relaxing the condition on perfect state preparation, we can obtain approximate fast control protocols which are tractable. It is natural to construct the variational basis for these approximate protocols in Krylov space. Then, problem of finding approximate control protocols is entirely equivalent to approximating the function $1/\omega$ by odd polynomials in $\omega$. All information about the system only enters through the spectral function of the operator $\partial_\lambda H$.

In this work we do this with a fixed Chebyshev polynomial approximation of $1/\omega$ in a finite window $[\zeta(\ell),1]$, where $\zeta(\ell)\sim \log(\ell)/\ell$ and $\ell$ is the order of the polynomial. Then we use an overall energy scale $\Omega(\ell)$, which rescales this interval to $[\mu(\ell),\Omega(\ell)]$ where $\mu(\ell)=\zeta(\ell)\Omega(\ell)$, as the single remaining variational parameter. We find that its asymptotic dependence on $\ell$ is determined by the scaling of the high-frequency tail of the spectral function. In particular, if these tails decay faster than exponentially with frequency, the universal protocol is guaranteed to improve with $\ell$ even in the thermodynamic limit. However, in generic situations, where these tails are expected to be exponential, we find that
the scale $\Omega(\ell)$ needs to grow linearly with $\ell$ so that the lower cutoff $\mu(\ell)$ saturates at a finite value. This saturation prevents an approximate CD protocol from continuously improving with $\ell$ in the thermodynamic limit.

These protocols perform comparably with variational approaches that require knowledge of the Hamiltonian beforehand. There are several further improvements to this method that might be made. For example, the high-frequency tail of the spectral function usually converges quickly with the system size, such that one could compute this efficiently for small systems and extrapolate to the thermodynamic limit. Then one can use this extrapolated spectral function as a weight function in the polynomial fitting of $1/\omega$ and additionally improve convergence of the approximate CD protocol \cite{finžgar2025counterdiabaticdrivingperformanceguarantees}. 

Interestingly, the results of this work show that one can search for optimal annealing paths amenable to local CD driving by minimizing high-frequency tails of the spectral function and hence minimizing high-frequency noise and dissipation. These paths should be most suitable for fast adiabatic state preparation. This finding
suggests {\sm the} existence of some fundamental connections between short- and long-time response of interacting systems and needs further exploration. Lastly, this method might be fruitfully applied to practical quantum annealing where the problem Hamiltonian, or a model for the disorder/noise, is not exactly known.

\textit{Note:} In the latter stages of this project, the authors became aware of another work in progress [\onlinecite{finžgar2025counterdiabaticdrivingperformanceguarantees}] which arrives at similar results via different means. These works have been submitted simultaneously.


\section*{Acknowledgments}
The authors thank Dries Sels for many beneficial discussions, as well as Hyeongjin Kim and John Martyn. This work was supported by the AFOSR Grant FA9550-21-1-0342 and the NSF Grant DMR-2412542. The exact code used is available online \footnote{See https://github.com/smorawetz/CD-extra-controls.git}.

\appendix

\section{Extra factor of $\omega$ in cost function} \label{appendix:omega-cost-func}

As described in the main text, the established variational principle involves minimizing the action $S_\lambda$ defined in Eq. \eqref{eqn:variational_action}. Writing this explicitly in terms of matrix elements and then employing the definition of the spectral function in Eq. \eqref{eqn:spectral-fn} gives

\begin{equation} \label{eqn:var_action}
    S_\lambda = \min_{\beta_k} \int d\omega \left(\sum_k \beta_k T_{2k}(\omega) + g(\omega/\mu) \right)^2 \Phi_\lambda(\omega)
\end{equation}

\begin{figure}
    \centering
    \includegraphics[width=\linewidth]{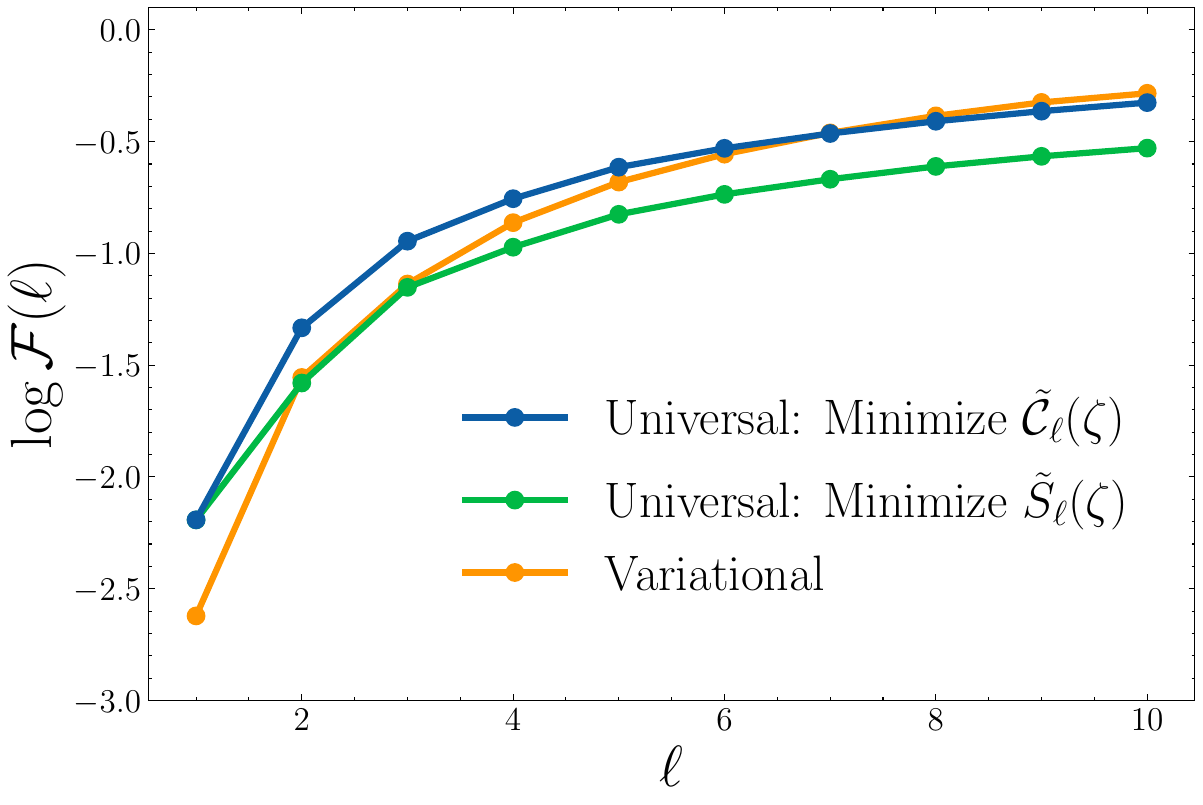}
    \caption{Comparison of performance of universal protocols based on the optimization of the cost function~\eqref{eqn:cost-fn-chebyshev}  and the action~\eqref{eqn:action-chebyshev}. We compare the final fidelity for the NNN TFI model annealing protocol when $N = 14$. There is very little difference between the two in terms of the final fidelity. Also plotted is the fidelity obtained by the variational approach, where despite having no knowledge of the Hamiltonian a priori, the fidelity difference is minimal.}
    \label{fig:compare-cost-funcs}
\end{figure}

This expression differs from the cost function $\mathcal{C}$ that is used in this work, defined in Eq. \eqref{eqn:cost-func}, by a factor of $\omega$ in the brackets. To assess the effect of this factor, we 
repeat the same analysis as in the main text by using the dimensionless action
\begin{equation} \label{eqn:action-chebyshev}
    \tilde{S}_\ell(\zeta) = \int_{\zeta}^1 dx \left(\sum_{k=1}^\ell \beta_k T_{2k} (x) + 1 \right)^2,
\end{equation}
instead of the cost function~\eqref{eqn:cost-fn-chebyshev} and repeat the same steps as in the main text. This procedure results in a  slightly different $\beta_k$ and $\zeta(\ell)$ due to modified integrand. We compare the final state fidelity for the NNN TFI model annealing protocol using this approach with that of the main text in Figure \ref{fig:compare-cost-funcs}, and find the difference is minimal.

\section{Constructing the AGP in terms of Chebyshev polynomials} \label{appendix:chebyshev-AGP}

The Chebyshev polynomials $T_n(x)$, where $x=\omega/\Omega$ are defined using the following recursive relation:
\begin{align*}
    T_0(x) & = 1 \\
    T_1(x) & = x \\
    T_{n+1}(x) & = 2 x T_n(x) - T_{n-1}(x)
\end{align*}

Then one can always uniquely represent 
\begin{equation} \label{eqn:appendix:cheby-agp}
    \sum_k \alpha_k x^{2k-1} = \sum_k \beta_k  T_{2k-1}(x)
\end{equation}

In the corresponding operator construction~\eqref{eqn:commutator_ansatz_tchebyshev} rescaling of $\omega\to x=\omega/\Omega$, is equivalent to rescaling the Hamiltonian appearing in the Liouvillian by $H \rightarrow H / \Omega$. With this, the coefficients $\beta_k$ then correspond to representing the AGP in terms of the operators $O_k$ defined as
\begin{align*}
    O_0 & = \partial_\lambda (H / \Omega) \\
    O_1 & = [(H / \Omega),  O_0] \\
    O_{n+1} & = 2 [(H/\Omega), O_n] - O_{n-1}
\end{align*}

in the following way:
\begin{equation}
    A_\lambda^{(\ell)} = \Omega \sum_{k=1}^\ell \beta_k O_{2k-1}
\end{equation}

We then obtain the $\beta_k$ by substituting picking the $\beta_k$ to minimize $\tilde{\mathcal{C}}_\ell$ in Eq. \eqref{eqn:cost-fn-chebyshev}.

\section{Numerical evidence of spectral function Gaussian decay in XXZ Model} \label{appendix:xxz-gaussian-decay}

The asymptotic growth of the so-called Lanczos coefficients $\{b_n\}$ is related to the high-frequency tail of the spectral function \cite{parkerUniversalOperatorGrowth2019} by the relation

\begin{equation} \label{eqn:appendix:lanc-coeff-growth}
    \lim_{n\to\infty} b_n \sim n^{1/\alpha} \iff \lim_{\omega\to\infty}\Phi(\omega) \sim \exp[-\Gamma\omega^\alpha]
\end{equation}

\noindent with the presence of logarithmic corrections in one dimension. Thus, in order to establish Gaussian decay of the spectral function, we can look at the asymptotic growth of the Lanczos coefficients.

Using a recently developed Julia package leveraging the Pauli string structure of spin-1/2 models \cite{loizeauQuantumManybodySimulations2024}, we can compute these Lanczos coefficients for very large systems. Although $\alpha = 2$ has been observed several times at the Heisenberg point, the growth at other nearby points has not been studied. We reparametrize the XXZ model as

\begin{equation} \label{eqn:appendix-xxz-param}
    H = -J (\sum_i \sigma_i^x \sigma_{i+1}^x \sigma_i^y \sigma_{i+1}^y) - \Delta (\sum_i \sigma_i^z \sigma_{i+1}^z)
\end{equation}

\begin{figure}
    \centering
    \includegraphics[width=\linewidth]{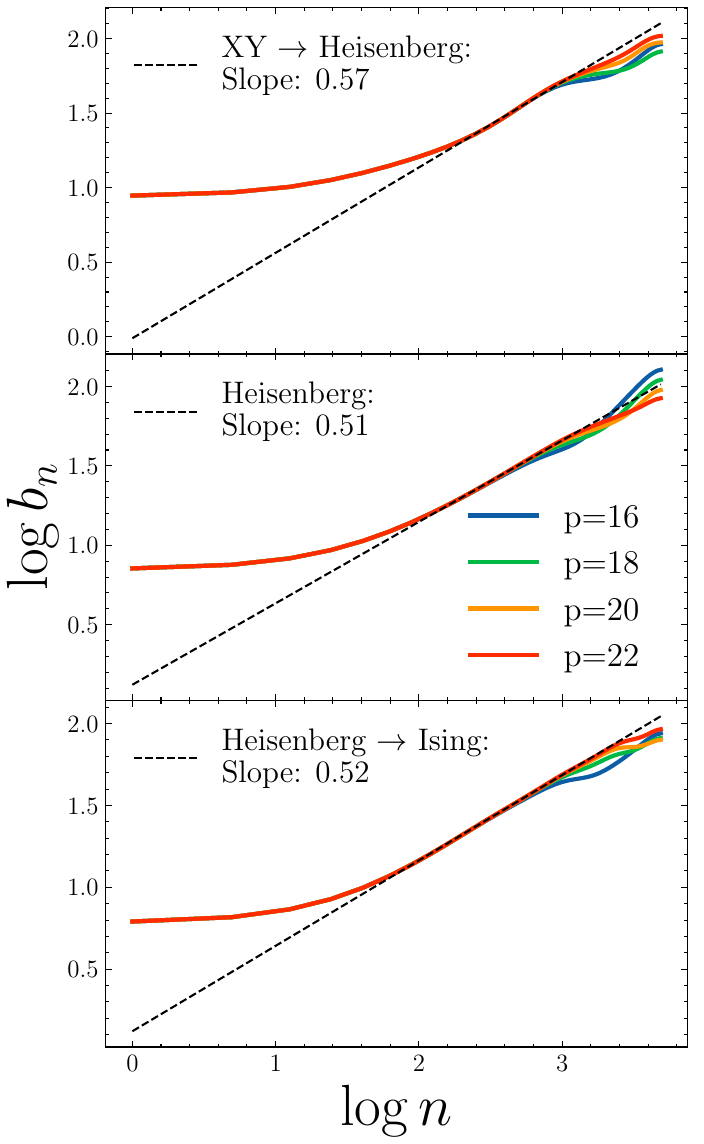}
    \caption{We show the growth of the first 40 Lanczos coefficients for system size $N = 50$ for our parametrization of the XXZ model at three different points. The numerics are reliable as long as $p = 22$ and $p = 20$ agree. We can see there is a well-defined regime of power law growth in each case, from which we extract the exponent $b_n \sim n^{1/\alpha}$. The first plot corresponds to $\Delta = J/2$ in Eq. \eqref{eqn:appendix-xxz-param}, the second is at the Heisenberg point ($\Delta = J$), and the third is at $\Delta = 2J$. In each instance, we average over a few nearby values of $\Delta$ and add a Gaussian filter to eliminate the high-frequency noise in the coefficients. The parameter $p$ corresponds to the maximum operator support in the calculation of the $b_n$, above which the components are truncated.}
    \label{fig:appendix:lanc-coeffs-XXZ}
\end{figure}

We will examine the growth of Lanczos coefficients for $\Delta = J/2$, $\Delta = J$ (the Heisenberg point) and $\Delta = 2J$. These are shown in Figure \ref{fig:appendix:lanc-coeffs-XXZ}. We see for $\Delta = J$ and $\Delta = 2J$ the growth agrees well with $\alpha = 2$. For $\Delta = J/2$ there is a noticeable deviation from $\alpha=2$, which could be a finite size effect. To avoid dealing with this, for the numerics in the main text, we choose to anneal from the Heisenberg point to a regime of strong anisotropy ($\Delta / J \to \infty$) as described.

\bibliography{bibliography}
\bibliographystyle{apsrev4-1}

\end{document}